# Thermal conductivity of solid parahydrogen with methane admixtures

A. I. Krivchikov[1], V. V. Sumarokov[2], J. Mucha[2], P. Stachowiak[2], A. Jeżowski[2], O. A. Korolyuk[1]

1 - B. Verkin Institute for Low Temperature Physics and Engineering, National Academy of Sciences of Ukraine, pr. Lenina 47, 61103 Kharkov, Ukraine

2 - W. Trzebiatowski Institute of Low Temperature Physics and Structure Research, Polish Academy of Sciences, P.O. Box 937, 50-950 Wroclaw, Poland

e-mail: krivchikov@ilt.kharkov.ua

The thermal conductivity of solid parahydrogen crystal with methane admixtures has been measured in the temperature range 1.5 to 8 K. Solid samples were grown from the gas mixture at 13 K. Concentration of $CH_4$ admixture molecules in the gas varied form 5 to 570 ppm. A very broad maximum of thermal conductivity with absolute value of about 110 W/(m·K) is observed at 2.6 K. The data are interpreted by Callaway model considering phonons resonant scattering on quasi-local vibrations of $CH_4$ molecules, phonon-grain boundary and phonon-phonon scattering processes. The increase of grain boundary scattering leads to the decrease of the maximum broadening. The analysis shows that the solid mixture of p-$H_2$ and $CH_4$ is a heterogeneous solution for $CH_4$ concentration higher than 0.1 ppm.

Crystalline parahydrogen can be used as an "inert" matrix to study the dynamics of the matrix-isolated atoms, molecules and their interaction with the crystalline surrounding [1-5]. Thermal conductivity is a property, which is very sensitive to the dynamic impurity – matrix interaction. At helium temperature ($T<\theta$) the inelastic anharmonic phonon-phonon $U$-processes in p-$H_2$ freeze out. A considerable part of its molecules are in the rotational ground state. As a result, the defect-free pure p-$H_2$ crystal has a very high thermal conductivity [6-9]. Introduction of even small amounts of molecular or atomic impurity decreases strongly the value and behavior of the temperature dependence of thermal conductivity $K(T)$ in the temperature region near the phonon maximum [9-11]. The main source of phonon scattering is the dynamic disorder in the crystal lattice caused by the dramatic mass difference between the substitutive impurity and the matrix molecule. Another factor reducing $K(T)$ may be the resonance phonon scattering at molecular centers – orthohydrogen clusters or impurity molecules with rotational degrees of freedom [12, 13].

In this study $K(T)$ of pure p-$H_2$ and $CH_4$-doped p-$H_2$ was measured to find out how the rotational motion of the $CH_4$ molecules affects $K(T)$ p-$H_2$. The $CH_4$ molecule in the $H_2$ matrix can be treated both as a heavy point defect (($m_{CH4}$ - $m_{H2}$)/$m_{H2}$ = 7) and as a weakly hindered quantum $CH_4$ rotor [1, 2]. Strong resonance phonon scattering by the rotational $CH_4$ excitations was detected earlier: it manifested itself as a dip in the temperature dependence of $K(T)$ for solid krypton with $CH_4$ impurity [14].

The experiment was carried out using a specially designed L-$^4$He cryostat [15]. The samples were prepared and $K(T)$ was measured directly in the sample chamber of the cryostat, in a glass ampoule (6.7 mm diameter and 67 mm long). The samples were grown from p-$H_2$ gas, at temperature near 13 K. After controlled cooling of the sample, down to 4.2 K, its $K(T)$ was measured by the steady state flow method, in the temperature range 1.2 ÷ 8 K. The samples were prepared using $H_2$ gas, 6.0 (99,9999 vol.%, Messer Co.) and $CH_4$ gas (99.95%). The chemical impurities in $H_2$ were $O_2 \leq 0.5$ vpm, $N_2 \leq 0.5$ vpm, $H_2O \leq 0.5$ vpm, HC $\leq 0.1$ vpm, CO/$CO_2 \leq 0.1$ vpm. The $CH_4$ gas contained 1.76% $CH_3D$, 0.12% $CHD_3$, 0.043% $N_2$, and 0.007% $O_2$. Parahydrogen containing less than 0.2% of orthohydrogen was obtained by conversion of L-$H_2$ contacting with Fe(OH)$_3$ below 20 K. The gas mixtures were prepared in a stainless steel vessel at room temperature. The error in the $CH_4$ concentration estimate in the mixture was below 20%. Optical polarized light observation showed that all the solid samples consisted of several (two or three) parts, which differed in the directions of the *c* – axis of their hcp lattice. Unlike pure p-H$_2$, the p-H$_2$-CH$_4$ samples were multicolored and their colors changed with the angle between the light polarization vector and the *c* – axis.

$K(T)$ was measured for several pure p-H$_2$ samples and (p-H$_2$)$_{(1-c)}$(CH$_4$)$_c$ solutions. The CH$_4$ concentration (*c*) in the initial gas mixture varied from 5 to 570 ppm. The experimental temperature dependences of $K(T)$ of pure p-H$_2$ for this work and literature data [7-9] are shown in Fig. 1. $K(T)$ of pure p-H$_2$ at temperature range of the phonon maximum agrees well with literature data [7, 9], but it is ten times lower than the record-high value reported in [8]. The good agreement observed for the thermal conductivity's of pure p-H$_2$ measured in three independent experiments may be due to the close isotopic (natural) compositions of the H$_2$ gas used in the this works. To put it differently, $K(T)$ at its maximum was limited by phonon scattering at H$_2$ isotopes, mainly HD molecules. The isotopic ratio R=[D]/[H] for H$_2$ of natural composition is $1.39 \div 1.56 \cdot 10^{-4}$. The mass-spectrometric analysis of gaseous H$_2$ showed that HD molecules were present in the initial gas. The considerable difference between the record-high $K(T)$ and the results of this study can be interpreted as an isotopic effect in $K(T)$ of p-H$_2$. The isotopic composition of H$_2$ was not specified in [8]. However, judging from the value of $K(T)$ maximum, the HD content in H$_2$ is an order of magnitude lower than in the gas of natural isotopic composition. At temperatures below the phonon maximum, $K(T)$ is influenced by the structural imperfections of the sample. Above the maximum it is influenced by *U*-processes which are in turn sensitive to the hcp lattice anisotropy of p-H$_2$ [16]. The contribution of the CH$_4$ impurity to $K(T)$ of solid p-H$_2$ can be detected only in case it exceeds the isotopic effect. Fig. 2 shows the calculated $K(T)$ curves for p-H$_2$ with different CH$_4$ concentrations (0.1, 1 and 10 ppm) assuming that the CH$_4$ molecules scatter phonons as heavy point defects.

The experimental dependences of $K(T)$ for pure p-H$_2$ and three p-H$_2$-CH$_4$ samples are shown in Fig. 3. $K(T)$ for the sample with 500 ppm CH$_4$ is only slightly lower than that for pure p-H$_2$ and is even higher than for the sample with 28 ppm CH$_4$. The CH$_4$ effect on $K(T)$ of p-H$_2$ is found to be weak. One of the reasons for this behavior may be very low solubility of CH$_4$ in solid H$_2$ (lower than 0.1 ppm). Researched samples consist in heterogeneous solid solutions. The main factors limiting the thermal conductivity are phonon scattering at the grain boundaries, HD impurity and the *U*-processes. Let's mark that spectroscopy of the matrix – isolated molecules in solid H$_2$ samples with 10 ppm grown at T≈8K detected no significant deviation of the CH$_4$ from random configuration distribution [1,2].

The experimental results were analyzed in the framework of Callaway theory allowing for the special role of normal phonon-phonon scattering processes in the thermal conductivity and using the Debye approximation to describe the phonon. In solid p-$H_2$ the resistive processes are formed by phonon-phonon processes $U$-processes ($\tau_U(x,T)$), scattering by the boundaries of crystalline grains and low-angle boundaries ($\tau_B(x,T)$), and the scattering caused by isotopic disorder (by the HD molecules) ($\tau_I(x,T)$): $\tau_R^{-1}(x,T) = \tau_U^{-1}(x,T) + \tau_B^{-1}(x,T) + \tau_I^{-1}(x,T)$. The characteristics of three-phonon $U$ and $N$ processes are determined only by the properties of the hcp lattice of $H_2$ and are practically independent of the impurity molecules at their low concentration: $\tau_U^{-1}(x,T) = A_U x^2 T^3 e^{(-E/T)}$, $\tau_N^{-1}(x,T) = A_N x^2 T^5$. The parameters $A_U$ and $E$ depend on the heat flow orientation with respect to the $c$-axis of the hcp lattice of $H_2$ [16]. The boundary scattering is dependent on the mean crystal grain size $L$: $\tau_B^{-1} = s/L$. The intensity of $N$-processes is taken from [17] ($A_N = 6.7 \cdot 10^5$ s$^{-1}$K$^{-5}$).

The scattering related to isotope disorder is characterized by the relaxation rate having the form of the Rayleigh formula: $\tau_I^{-1} = \xi \left(\dfrac{\Delta m}{m_{H2}}\right)^2 \dfrac{\Omega_0}{4\pi s^3} \omega^4$, where $\xi$ is the HD concentration, $\Omega_0$ is the volume per atom, $\Delta m = 1$ for HD impurity in $H_2$. The fitting parameters used to match the curves calculated and experimental results were the HD concentration $\xi$ and the parameters $L, A_U, E$ (see Table). It was assumed for p-$H_2$-$CH_4$ samples that the concentration $\xi$ was invariable from sample to sample. Parameter $L$ characterises the structure of defects in the sample [18]. $L$ varied nearly fivefold (see Table). This variation of $L$ in the samples prepared from the gas mixtures p-$H_2$-$CH_4$ indicates that the density of low-angle boundaries in the sample increases due to the stress appearing during the growth of the crystal and on its cooling.

The new measurement result for $K(T)$ of pure p-$H_2$ has been interpreted as a manifestation of the isotopic effect. $K(T)$ of the samples prepared by depositing gas mixtures (p-$H_2$)$_{(1-c)}$($CH_4$)$_c$ in the interval 5 – 570 ppm at temperatures near the triple point of p-$H_2$ varied only slightly with concentration. Because of weak solubility of $CH_4$ in p-$H_2$, we were unable (unlike in the Kr-$CH_4$ case) to detect resonance phonon scattering by rotational excitations of the $CH_4$ molecules (a dip in the temperature dependence of $K(T)$ of solid $H_2$ with $CH_4$ impurity). The upper limit of $CH_4$ solubility in solid p-$H_2$ was estimated from thermal conductivity values. It does not exceed 0.1 ppm.

We are grateful to Prof. V.G. Manzhelii, Dr. B. Ya. Gorodilov for helpful discussions. The study described in this publication was made possible in part Grant N 2M/78-2000 of Ukrainian Min. of Education and Science.

Table. Best fitting parameters, phonon mean free path $L$ (boundary scattering), HD concentration $\xi$, parameters $A_U$, $E$, were found from the analysis of experimental results $K(T)$ for different samples of pure p-H$_2$ and p-H$_2$-CH$_4$.

| Sample | $\xi$, ppm | $L$, mm | $A_U$, K$^{-3}$ s$^{-1}$ | $E$, K |
|---|---|---|---|---|
| p-H$_2$ sample 1 | 300±40 | 0.57±0.05 | 4.20·10$^7$ | 36 |
| p-H$_2$ sample 2 | 300±40 | 1.11±0.05 | 2.11·10$^7$ | 36 |
| p-H$_2$+28 pmm CH$_4$ | 300±40 | 0.34±0.05 | 4.80·10$^7$ | 36 |
| p-H$_2$+116 pmm CH$_4$ | 300±40 | 0.17±0.05 | 4.20·10$^7$ | 36 |
| p-H$_2$+500 pmm CH$_4$ | 300±40 | 0.71±0.05 | 4.20·10$^7$ | 36 |

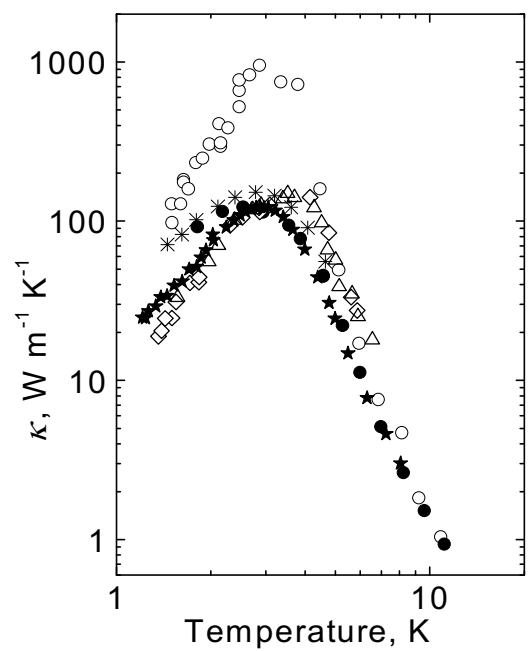

Fig. 1. The temperature dependence of thermal conductivity of solid pure parahydrogen.
○ - [8]; ● - [9]; ◊, Δ - [7] for p-$H_2$ with 0.2% o-$H_2$ and p-$H_2$ with 0.34% o-$H_2$ accordingly; ★, ∗ - present data for sample 1, 2.

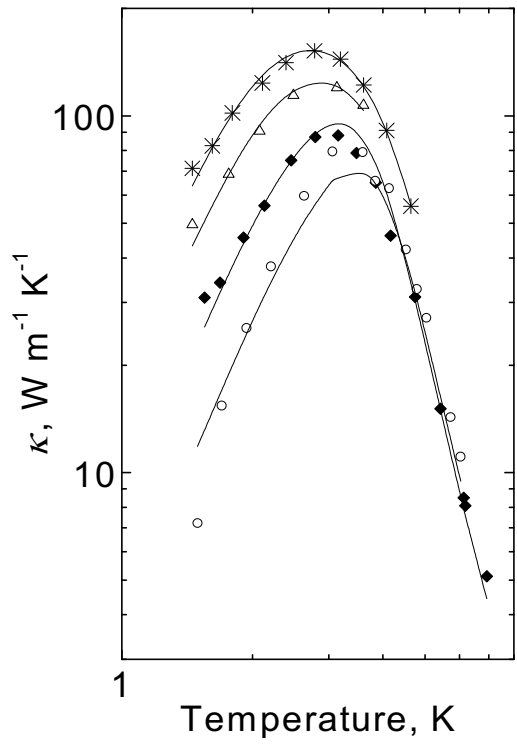

Fig. 3. The temperature dependence of thermal conductivity of $(p-H_2)_{(1-c)}(CH_4)_c$. Points are experimental dates: ∗ - the pure parahydrogen, sample 2; ◆ - $c = 28$ ppm $CH_4$; ○ - $c = 116$ ppm $CH_4$; Δ - $c = 500$ ppm $CH_4$. Solid lines are the best fitting curves.

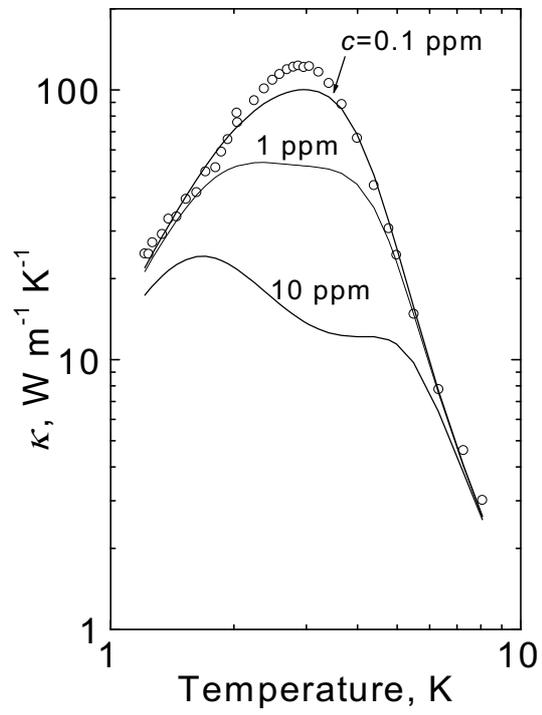

Fig. 2. The temperature dependence of thermal conductivity of parahydrogen with $CH_4$ admixture. Points are experimental data of this work for the pure parahydrogen, solid lines are the theoretical calculations for $c$ = 0.1, 1, 10 ppm $CH_4$ in p-$H_2$.